\documentclass[prd,twoside,twocolumn,longbibliography]{revtex4-2}
\usepackage{amsmath}
\usepackage{amssymb}
\usepackage{amsfonts}

\renewcommand{\vec}{\textbf}
\newcommand{\ket}[1]{|#1\rangle}
\newcommand{\bra}[1]{\langle#1|}
\newcommand{\bracket}[2]{\langle#1|#2\rangle}

\newcommand{\fb}[1]{\mathsf{B}(#1)}
\newcommand{\id}{{\mathbb{I}}}
\DeclareMathOperator{\tr}{Tr} 
\usepackage{subfig}
\usepackage{graphicx}
\usepackage{color}
\begin{document}

\title{Entanglement and Bell inequalities violation in $H\to ZZ$ with anomalous coupling}

\author{Alexander Bernal}
\email{alexander.bernal@csic.es}
\affiliation{Instituto de F\'isica Te\'orica, IFT-UAM/CSIC, 
	Universidad Aut\'onoma de Madrid,\\
	Cantoblanco, 28049 Madrid, Spain}
\author{Pawe{\l}{} Caban}
\email{Pawel.Caban@uni.lodz.pl (corresponding author)}
\author{Jakub Rembieli\'nski}
\email{jaremb@uni.lodz.pl}
\affiliation{Department of Theoretical Physics,
	University of {\L}{\'o}d{\'z}\\
	Pomorska 149/153, PL-90-236 {\L}{\'o}d{\'z}, Poland}

\date{\today}

\begin{abstract}
We discuss entanglement and violation of Bell-type inequalities for a system
of two $Z$ bosons produced in Higgs decays. We take into account beyond
the Standard Model (anomalous) coupling between $H$ and daughter bosons
but we limit ourselves to an overall scalar $ZZ$ state (we exclude the possibility
that $H$ contains a pseudo-scalar component). In particular we consider
the case when each $Z$ decays further into fermion-antifermion pair.
We find that the $ZZ$ state is entangled and violates the CGLMP inequality for
all values of the (anomalous) coupling constant.
We also discuss the impact of a background on these results.
The methods we develop are completely general, since they
can be extrapolated to any scalar particle decaying into two spin-1 particles 
of different masses. Moreover, the violation of the CGLMP inequality in the final state 
is theoretically ensured for any value of the couplings.
\end{abstract}

\maketitle

\section{Introduction}

Violation of Bell inequalities is one of the most striking properties of quantum theory.
Such a violation has been observed in a variety of physical systems like e.g. pairs of photons
\cite{FC1972,ADR1982,GVetal2015}, ions \cite{cab_RKMSIMW2001}, 
electrons \cite{HBDetal2015}, superconducting currents \cite{AWetal2009_josephson}
or solid state systems \cite{PTetal2013}.

Recently, the possibility of observing quantum entanglement and
violation of Bell-type inequalities
in high energy physics has been put forward.
In particular, 
scattering processes \cite{SZ2023_Bell_2-2-scattering,Morales2023-boson-scattering},
systems of top quarks \cite{AN2021-top-quarks,FFP2021-PhysRevLett.127.161801,ASC2022_top-quarks,AMMM2022_tt-quarks,AguilarSaavedra2023_postdecay,DGKN2023-Bell-entanglement-tt},
$B^0\bar{B}^0$ mesons \cite{TIetal2021-PhysRevD.104.056004}
and pairs of vector bosons arising from Higgs particle decay 
\cite{Barr2021,BCR2022-Bell-vector-bosons,ASBCM2022_entanglement_HtoZZ,APBW2022_entanglement-weak-decays,Aguilar-Saavedra2022_entanglement_HtoWW,FFGM2023}
were proposed in this context.
This last possibility for the first time was suggested by Alan Barr in \cite{Barr2021}.
Barr considered there the violation of Clauser--Horn--Shimony--Holt (CHSH) and
Collins--Gisin--Linden--Massar--Popescu (CGLMP) ine\-qualities in a system of 
$WW$ bosons arising in the decay of Higgs particle.
In \cite{BCR2022-Bell-vector-bosons} the possible violation of CHSH, Mermin
and CGLMP inequalities for a boson--antiboson system in an overall scalar state was 
discussed.
In this paper the most general scalar state of two boson system was considered.
If the bosons originate from the Higgs decay, then one of the components
of such a general scalar state corresponds to an anomalous coupling of $H$ with 
the daughter bosons (we explain this point in the present paper in
Sec.~\ref{sec:state-Higgs-decay}). Such a general scalar state of two vector
bosons in the context of quantum correlations
for the first time was discussed in \cite{Caban_2008_bosons_helicity}
while the correlations of relativistic vector bosons in \cite{CRW_2008_vector_bosons}.
The authors of \cite{ASBCM2022_entanglement_HtoZZ} analyzed entanglement
and violation of CGLMP inequality in the system of two $Z$ bosons produced in the decay
of a Higgs particle assuming the Standard Model interaction of $H$ with the $ZZ$ pair.
In \cite{Aguilar-Saavedra2022_entanglement_HtoWW} entanglement of $W$
bosons produced in $H\to WW \to l\nu l\nu$ channel was considered.
The paper \cite{APBW2022_entanglement-weak-decays} explores the possibility of
using quantum state tomography methods to determine a density matrix of massive
particles produced in weak decays.
In \cite{FFGM2023} entanglement and Bell inequality violation in boson pairs arising
in the Standard Model processes $H\to W W^*$, $H\to Z Z^*$, $pp\to WW$,
$pp\to WZ$, and $pp\to ZZ$ is considered.

In this paper we discuss entanglement and violation of the CGLMP inequality
for a $ZZ$ system produced in Higgs decay. We consider anomalous (beyond
the Standard Model) structure of the vertex describing interaction of a Higgs
particle with two daughter bosons but limit ourselves to the case
of a scalar Higgs.
Anomalous coupling parameters in the amplitude describing the interaction
between $H$ and $ZZ$ bosons are strongly constrained by measurements of Higgs
properties performed at the LHC
\cite{CMSCollab2019-H-anomalous-PhysRevD.99.112003},
we discuss this point in detail in next section, below Eq.~(\ref{c-v_2-v_1}).
However, our analysis is based on the most general Lorentz-covariant,
CPT conserving amplitude (Eq.~(\ref{general-vertex})) describing coupling
of a (pseudo)scalar particle with two spin-1 particles with different masses.
Thus, our considerations can be applied to each of such processes.
The Higgs decay can be treated as an exemplary process of this kind.

It is also worth to notice the very recent papers 
\cite{FFGM2023_anomalous,AMMM2023-New-phys-diboson}.
The authors of the former paper propose to use quantum tomography techniques to bound
anomalous coupling in $H\to WW$ and $H\to ZZ$ decays while the authors of the latter one
use entanglement to probe new physics in diboson production.

We use the standard units ($\hbar=c=1$, here $c$ stands for the velocity of light) 
and the Minkowski metric tensor 
$\eta=\mathrm{diag}(1,-1,-1,-1)$.

\section{State of the two-boson system arising from the Higgs decay}
\label{sec:state-Higgs-decay}

Our first goal is to construct a quantum state of two bosons arising in the process
\begin{equation}
H\to ZZ.
\label{HZZ}
\end{equation}
Let us denote by $M$ the Higgs mass and by $k,m_1$ and $p,m_2$ the four-momenta
and invariant masses of the in general off-shell $Z$ bosons produced in the decay (\ref{HZZ}).
In the actual decay of the Higgs particle into a pair of $Z$ bosons typically one of 
them is nearly on-shell and the other one off-shell. Nevertheless, for the sake of generality,
we consider $Z$ bosons with arbitrary invariant masses. We decided to work in this
framework as it covers all possible scenarios of a general on-shell scalar decaying
into vector bosons. On the other hand, when we average the $ZZ$ state over kinematical 
configurations we use the probability distribution $\mathcal{P}_c(m_1,m_2)$
(Eq.~(\ref{rho_ZZ_mixed-def})) giving the probability that $H$ decays into bosons
with masses $m_1$ and $m_2$. For the actual process (\ref{HZZ}) this probability density is 
peaked at $m_i=m_Z$ ($i=1$ or 2). For further remarks on this point see the paragraph
above Eq.~(\ref{rho_ZZ-with-noise}).
Moreover, we treat of-shell particles like ordinary on-shell particles with reduced masses.
Similar approach has been applied in previous quantum-information-related studies
\cite{ASBCM2022_entanglement_HtoZZ,FFGM2023} as well as more 
phenomenologically-oriented papers like \cite{GMM_2007-HZZ,ZK_2016}. 

We will perform our computations in the center of mass (CM) frame.
In this frame we denote energies of $Z$ bosons as $\omega_1$ and
$\omega_2$, consequently $k^\mu=(\omega_1,\vec{k})$,
$p^\mu=(\omega_2,-\vec{k})$ and 
$\omega_1^2 - \vec{k}^2 = m_1^2$, $\omega_2^2 - \vec{k}^2 = m_2^2$.

Using similar notation as in
\cite{BCR2022-Bell-vector-bosons,CRW_2008_vector_bosons,Caban_2008_bosons_helicity},
a general scalar state of two vector bosons with arbitrary masses can be written as
\begin{equation}
	\ket{\psi_{ZZ}^{\mathsf{scalar}}(k,p)} = 
	g_{\mu\nu}(k,p) 
	e_{\lambda}^{\mu}(k) e_{\sigma}^{\nu}(p)
	\ket{(k,\lambda);(p,\sigma)},
	\label{scalar-state-general}
\end{equation}
where
\begin{equation}
	\label{g_def}
	g_{\mu\nu}(k,p) =  \eta_{\mu\nu} 
	+ \tfrac{c}{(kp)} \big( k_\mu p_\nu + p_\mu k_\nu \big),
	\quad c \in{\mathbb{R}},
\end{equation}
and $\ket{(k,\lambda);(p,\sigma)}$ denotes the two-boson state, one boson with 
the four-momentum $k$ and spin projection along $z$ axis $\lambda$,
second one with the four-momentum $p$ and spin projection $\sigma$.
The basis two-particle states fulfill the following orthogonality condition
(for $k\not= p$)\footnote{Here, for convenience, we use vectors which are rescaled 
with respect to the basis vectors used in \cite{CRW_2008_vector_bosons} or
\cite{BCR2022-Bell-vector-bosons}. To obtain vectors used here one have to multiply 
those from \cite{BCR2022-Bell-vector-bosons} by 
$(2\delta^3(\vec{0})\sqrt{\omega_1 \omega_2})^{-1}$.}:
\begin{equation}
	\bracket{(k,\lambda);(p,\sigma)}{(k,\lambda^\prime);(p,\sigma^\prime)} 
	=  \delta_{\lambda\lambda^\prime} 
	\delta_{\sigma \sigma^\prime}.
	\label{normalization-basis}
\end{equation}

The explicit form of amplitude $e_{\lambda}^\mu(q)$ for
the four-momentum $q=(q^0,\vec{q})$ with ${q^0}^2-{\vec{q}}^2=m^2$
reads \cite{CRW_2008_vector_bosons}
\begin{equation}
	e(q) = [e^{\mu}_{\sigma}(q)] = 
	\begin{pmatrix}
		\tfrac{\vec{q}^T}{m}\\
		\id + \tfrac{\vec{q}\otimes \vec{q}^T}{m(m+q^0)}
	\end{pmatrix}
	V^T, 
	\label{amplitude-e-explicit}
\end{equation}
and
\begin{equation}
	\label{matrix_V}
	V=\frac{1}{\sqrt{2}}
	\begin{pmatrix}
		-1 & i & 0 \\
		0 & 0 & \sqrt{2} \\
		1 & i & 0 \\
	\end{pmatrix}.
\end{equation}
These amplitudes fulfill standard transversality condition
\begin{equation}
e^{\mu}_{\sigma}(q) q_\mu = 0.
\label{transversality}
\end{equation}

To find an interpretation of the parameter $c$ introduced in
Eqs.~(\ref{scalar-state-general},\ref{g_def}) let us notice that
the two boson state can be computed using the structure of the vertex
describing interaction of the Higgs particle with two daughter vector bosons.
Following e.g. \cite{ZK_2016,GMM_2007-HZZ} the amplitude corresponding to the 
most general Lorentz-invariant, CPT conserving coupling of the (pseudo)sca\-lar 
particle with two vector bosons can be cast in the following form:
\begin{multline}
	\mathcal{A}_{\lambda\sigma}(k,p) \propto
	\big[v_1 \eta_{\mu\nu} + v_2 (k+p)_\mu (k+p)_\nu \\
	+v_3 \varepsilon_{\alpha\beta\mu\nu} (k+p)^\alpha (k-p)^\beta \big]
	e_{\lambda}^{\mu}(k) e_{\sigma}^{\nu}(p),
	\label{general-vertex}
\end{multline}
where $\lambda,\sigma$ are spin projections of the final states, 
$v_1$, $v_2$, $v_3$ are three real coupling constants, and
$\varepsilon_{\alpha\beta\mu\nu}$ is a completely antisymmetric 
Levi-Civita tensor.

The Standard Model interaction corresponds to $v_1=1$, $v_2=v_3=0$.
On the other hand, $v_3\not=0$ implies that Higgs boson contains a pseudo-scalar
component and indicates the possibility of CP violation.
For the moment, let us consider the case $v_3=0$, $v_1\not=0$, and $v_2$ free.
Comparing (\ref{scalar-state-general},\ref{g_def}) with (\ref{general-vertex}) 
and taking into account the transversality condition (\ref{transversality})
we can relate the parameter $c$ with 
the coupling constants $v_1$, $v_2$
\begin{equation}
	c= \frac{v_2}{v_1} (kp).
	\label{c-v_2-v_1}
\end{equation}
Let us mention that the assumption $v_1=0$, apart from being unphysical
(we know that the Higgs has $v_1\not=0$), together with $v_3=0$ leads 
to a separable state.
Therefore, from now on we limit ourselves to the case 
$v_3=0$, $v_1\not=0$, $v_2$ free,
that is we assume that the Higgs boson is a scalar but we admit a beyond Standard 
Model coupling $v_2\not=0$.

We would like to stress here that there exist experimental bounds
on anomalous couplings $v_2$ and $v_3$. Strong bound comes from the 
measurements of Higgs boson particles performed at the LHC by the CMS
Collaboration \cite{CMSCollab2019-H-anomalous-PhysRevD.99.112003}.
The CMS Collaboration paper uses a different parametrization of the amplitude 
describing the interaction between $H$ and two daughter bosons,
instead of $v_1$, $v_2$ they use parameters $a_{1}^{ZZ}$ and $a_{2}^{ZZ}$---see
Eq.~(2) in \cite{CMSCollab2019-H-anomalous-PhysRevD.99.112003}.
Comparing (2) in \cite{CMSCollab2019-H-anomalous-PhysRevD.99.112003}
and our Eq.~(\ref{general-vertex}) we obtain the following relation between
the parameterizations:
\begin{equation}
v_1 \propto a_{1}^{ZZ} m_{Z}^{2} + 2 a_{2}^{ZZ} (kp),\quad
v_2 \propto -2a_{2}^{ZZ}
\end{equation}
(with the same proportionality constant).
Thus, following (\ref{c-v_2-v_1})
\begin{align}
c & = 2 \left( \tfrac{a_{2}^{ZZ}}{a_{1}^{ZZ}} \tfrac{(kp)}{m_{Z}^{2}}\right)
\tfrac{-1}{1 + 2 \left( \tfrac{a_{2}^{ZZ}}{a_{1}^{ZZ}}
\tfrac{(kp)}{m_{Z}^{2}}\right)} \,\,\Rightarrow \nonumber\\
|c| & = 2 \left| \tfrac{a_{2}^{ZZ}}{a_{1}^{ZZ}}\right| \tfrac{(kp)}{m_{Z}^{2}}
+ \mathcal{O}\Big(\left| \tfrac{a_{2}^{ZZ}}{a_{1}^{ZZ}}\right|^2\Big).
\label{c_exp-bound-1}
\end{align}
Now, the experimental bounds on the ratio $a_{2}^{ZZ}/a_{1}^{ZZ}$ are given in
Table 7 of \cite{CMSCollab2019-H-anomalous-PhysRevD.99.112003} and in the 
on-shell case they read:
$a_{2}^{ZZ}/a_{1}^{ZZ}\in [-0.12,0.26]$ at 95\% C.L.
Thus, taking a larger value in this range, i.e. 
$|a_{2}^{ZZ}/a_{1}^{ZZ}|< 0.26$, and neglecting terms of order
$|a_{2}^{ZZ}/a_{1}^{ZZ}|^2$ and higher, we get
\begin{equation}
|c| < 0.26 \tfrac{2(kp)}{m_{Z}^{2}}.
\label{c_exp-bound-2}
\end{equation}
To estimate the maximal value of $2(kp)/m_{Z}^{2}$ we use Eq.~(\ref{formula_2})
and assume that one of the $Z$ bosons is on-shell and the invariant mass of the off-shell
$Z$ boson is equal to zero. 
Inserting the measured values for the Higgs mass $M=125.25\; \mathsf{GeV}$
and $Z$ mass $m_Z=91.19\; \mathsf{GeV}$ \cite{Workman:2022-PDG} we 
finally obtain the following bound for experimentally admissible values of $c$ in the
process $H\to ZZ$:
\begin{equation}
|c| < c_{\mathsf{HZZ}}^{\mathsf{max}} = 0.23.
\label{cHZZmax}
\end{equation}
Ref.~\cite{FFGM2023_anomalous} suggests that even stronger bound could 
be obtained using the tomography of the two-boson density matrix.
Therefore, in the actual process $H\to ZZ$ the range of the parameter $c$ which 
is not excluded by experimental data is rather narrow.
However, as we mentioned in Introduction, the process $H\to ZZ$ can be treated
as a model for the most general case of a decay of a (pseudo)scalar particle into
two spin-1 particles with different masses. That is why in the following part of the 
paper we do not restrict values of $c$ to the interval
$(-c_{\mathsf{HZZ}}^{\mathsf{max}},c_{\mathsf{HZZ}}^{\mathsf{max}})$.

The following comment is also in order here: when we consider 
a decay of a scalar particle into two gauge bosons, gauge invariance requires that 
$v_2$ coupling appears in the combination $v_2(k_\mu p_\nu - \eta_{\mu\nu}(kp))$
which is equivalent to $c=-1$. Thus, the cases $c=0$ and $c=-1$ are indeed special
as it was emphasized in \cite{BCR2022-Bell-vector-bosons}.

The normalization of the scalar state defined in
Eq.~(\ref{scalar-state-general}) is the following:
\begin{equation}
	\bracket{\psi_{ZZ}^{\mathsf{scalar}}(k,p)}{\psi_{ZZ}^{\mathsf{scalar}}(k,p)} 
	=  A(k,p),
	\label{normalization-scalar}	
\end{equation}
with
\begin{equation}
	A(k,p) = 2  +
	\Big[ (1+c)\tfrac{(kp)}{m_1 m_2} - c \tfrac{m_1 m_2}{(kp)} \Big]^2
	\equiv 2 + \kappa^2,
	\label{A_k-p}
\end{equation}
where, for further convenience, we have introduced the parameter $\kappa$.
Using formulas (\ref{energy_conservation}-\ref{formula_4}) we find that in the CM frame
$\kappa$ depends only on masses $M$, $m_1$, $m_2$ and the parameter $c$:
\begin{equation}
\kappa = \beta + c (\beta - 1/\beta),
\label{kappa}
\end{equation}
where
\begin{equation}
	\beta = \frac{M^2 - (m_1^2+m_2^2)}{2m_1 m_2}.
	\label{K-CM}
\end{equation}
The range of possible values of $\kappa$ depends on the value of $c$ and is the following:
\begin{align}
	&\kappa \in (-\infty,1] && \textrm{for} && c\in(-\infty,-1),\\
	&\kappa \in [0,1] && \textrm{for} && c=-1,\\
	&\kappa \in [2\sqrt{-c(1+c)},\infty) && \textrm{for} && c\in(-1,-\tfrac{1}{2}),\\
	&\kappa \in [1,\infty] && \textrm{for} && c\in[-\tfrac{1}{2},\infty).
\end{align}
Nevertheless, further theoretical constraints must be taken into
account to give the physically allowed range for $c$. 
In particular, perturbative unitary 
(see \cite{Logan2022_Perturbative-unitarity-Higgs} for a recent review 
applied to Higgs physics) imposes bounds over the values of the anomalous
coupling $v_2$. 
Namely, based on $ZZ\to ZZ$ scatterings,
numerical bounds have been obtained for the $H\to ZZ$ anomalous couplings
\cite{DDI2016_Perturbative-unitarity-anomalous}. 
Comparing our amplitude (Eq. (\ref{general-vertex})) with Eq.~(9) from
\cite{DDI2016_Perturbative-unitarity-anomalous}
we get the following relation between our parametrization and  pa\-ra\-me\-trization
used in \cite{DDI2016_Perturbative-unitarity-anomalous}:
\begin{equation}
v_1 \propto a^{ZZH}_1 m_Z^2 - a^{ZZH}_2 (kp),\quad 
v_2 \propto a^{ZZH}_2.
\end{equation}
Thus, following (\ref{c-v_2-v_1})
\begin{equation}
c  =  \left( \tfrac{a_{2}^{ZZH}}{a_{1}^{ZZH}} \tfrac{(kp)}{m_{Z}^{2}}\right)
\tfrac{1}{1 -  \left( \tfrac{a_{2}^{ZZH}}{a_{1}^{ZZH}}
\tfrac{(kp)}{m_{Z}^{2}}\right)} .
\label{c_unitarity-bound-1}
\end{equation}
Now, we use Eq.~(\ref{formula_2})
assuming that one of the $Z$ bosons is on-shell and the invariant mass of the off-shell
$Z$ boson is equal to zero and apply
Eqs.~(11,12,25) from \cite{DDI2016_Perturbative-unitarity-anomalous}
to obtain 
\begin{equation}
\frac{(kp)}{m_{Z}^{2} a_{1}^{ZZH}} \approxeq 0.68158
\end{equation}
which gives
\begin{equation}
c \approxeq \frac{0.68158 a_{2}^{ZZH}}{1-0.68158 a_{2}^{ZZH}}.
\end{equation}
Therefore, taking into account that according to Table~I in
\cite{DDI2016_Perturbative-unitarity-anomalous}:
$|a^{ZZH}_2|<1.97$, we notice that $c$ has
a pole in the allowed range for $a^{ZZH}_{2}$ 
and hence $c\in(-\infty,\infty)$. That is, we
find that the requirement of perturbative unitarity does
not limit accessible values of $c$ in the process $H\to ZZ$.

Now, denoting
\begin{equation}
\vec{n} = \frac{\vec{k}}{|\vec{k}|},
\label{n_def}
\end{equation}
in the CM frame we can write the $ZZ$  scalar state (\ref{scalar-state-general}) as
\begin{equation}
\ket{\psi_{ZZ}^{\mathsf{scalar}}(m_1,m_2,\vec{n},c)} = 
\sum_{\lambda\sigma} \Omega_{\lambda\sigma} 
\ket{(k,\lambda);(p,\sigma)},
\label{scalar-ZZ-n}
\end{equation}
where $k^\mu=(\omega_1,0,0,|\vec{k}|\vec{n})$, 
$p^\mu=(\omega_2,0,0,-|\vec{k}|\vec{n})$, $\omega_1$, $\omega_2$, 
$|\vec{k}|$ are given by Eqs.~(\ref{formula_3},\ref{formula_4},\ref{formula_1}),
respectively, and
\begin{equation}
	\Omega = - \frac{1}{\sqrt{2+\kappa^2}} V (I+(\kappa-1)
	\vec{n}\otimes\vec{n}^T) V^T,
\end{equation}
($V$ is defined in Eq.~(\ref{matrix_V})).

Without loss of generality we take $\vec{n}$ along $z$ axis, $\vec{n}=(0,0,1)$
and simplify our notation:
\begin{equation}
\ket{((\omega_1,0,0,|\vec{k}|),\lambda);((\omega_2,0,0,-|\vec{k}|),\sigma)} 
\equiv \ket{\lambda,\sigma}.
\end{equation}
With such a choice we have
\begin{equation}
	\Omega = \frac{1}{\sqrt{2+\kappa^2}} 
	\begin{pmatrix}
		0 & 0 & 1 \\
		0 & -\kappa & 0 \\
		1 & 0 & 0
	\end{pmatrix},
\label{Omega-psiZZ}
\end{equation}
and the explicit form of the $ZZ$ state in this case reads
\begin{multline}
\ket{\psi_{ZZ}^{\mathsf{scalar}}(m_1,m_2,c)} = \frac{1}{\sqrt{2+\kappa^2}}
\Big[
\ket{+,-}\\
 - \kappa \ket{0,0} + \ket{-,+}
\Big].
\label{ZZ-state-CM}
\end{multline}
It is worth taking note that the form of the above state agrees with 
the most general form of the $ZZ$ state arising in the $H\to ZZ$ process conserving
parity---compare Eq.~(7) from \cite{ASBCM2022_entanglement_HtoZZ}.
Moreover, when we put $c=0$ in Eq.~(\ref{kappa}) the state (\ref{ZZ-state-CM})
coincides with the $ZZ$ state discussed in \cite{ASBCM2022_entanglement_HtoZZ}
where only the Standard Model vertex has been considered.

\subsection{Averaging over kinematical configurations}

In the realistic case when the state is determined on the basis of 
data obtained from various kinematical configurations, one has to 
average over these configuration. Thus, in such a case we receive a mixed state
which for a given $c$ reads
\begin{equation}
	\rho_{ZZ}(c) = \int dm_1\, dm_2\, \mathcal{P}_{c}(m_1,m_2) \rho(m_1,m_2,c),
	\label{rho_ZZ_mixed-def}
\end{equation}
where $\rho(m_1,m_2,c)=
\ket{\psi_{ZZ}^{\mathsf{scalar}}(m_1,m_2,c)}\bra{\psi_{ZZ}^{\mathsf{scalar}}(m_1,m_2,c)}$ (c.f.~(\ref{ZZ-state-CM})) and
$\mathcal{P}_{c}(m_1,m_2)$ is a normalized probability distribution.
To determine the form of $\mathcal{P}_{c}(m_1,m_2)$, we assume that each of the
$Z$ bosons produced in the process (\ref{HZZ}) decays subsequently into massless
fermion--antifermion pair, i.e. we consider the process
\begin{equation}
	H \to ZZ \to f_{1}^{+} f_{1}^{-} f_{2}^{+} f_{2}^{-}.
\end{equation}
The normalized differential cross section of the decay 
$ZZ \to f_{1}^{+} f_{1}^{-} f_{2}^{+} f_{2}^{-}$ 
is given by
\begin{equation}
	\frac{1}{\sigma} \frac{\sigma}{d\Omega_1\, d\Omega_2}
	=\Big(\frac{3}{4\pi}\Big)^2 
	\tr\big[ \rho_{ZZ}(c) (\Gamma_{1}^T \otimes \Gamma_{2}^T)\big],
	\label{cross-sec-decay-matrices}
\end{equation}
where $\Gamma_i$ is the decay matrix of $Z_i\to f_{i}^{+} f_{i}^{-}$, 
$\Omega_i$ is the solid angle related to the final particle $f_{i}$ \cite{RS2022-Decay-matrices,ASBCM2022_entanglement_HtoZZ}.
Now, inserting (\ref{rho_ZZ_mixed-def}) into (\ref{cross-sec-decay-matrices}),
integrating with respect to solid angles,
using the property that $\int \Gamma_i d\Omega_i = \tfrac{4\pi}{3} I$, and
differentiating with respect to $m_1$, $m_2$
we obtain
\begin{multline}
	\frac{1}{\sigma} \frac{d\sigma}{dm_1\, dm_2}(m_1, m_2, c)\\
	= \frac{\mathcal{P}_{c}(m_1,m_2)}{\tr[\tilde{\rho}(m_1,m_2,c)]}
	\tr[\tilde{\rho}(m_1,m_2,c)],
	\label{cross-sec-masses}
\end{multline}
where by $\tilde{\rho}(m_1,m_2,c)$ we have denoted non-normalized
density matrix $\rho(m_1,m_2,c)$, i.e.,
\begin{multline}
\tilde{\rho}(m_1,m_2,c)=
\Big[ \ket{+,-} - \kappa \ket{0,0} 
+ \ket{-,+}\Big] \times \\
\Big[ \bra{+,-} - \kappa \bra{0,0} + \bra{-,+}\Big].
\label{rho-non-normalized}
\end{multline}
Next, Eq.~(8) from \cite{ZK_2016} in our notation can 
be written as\footnote{Zagoskin and Korchin in \cite{ZK_2016} work 
	in the helicity basis. Note that with our choice of the reference frame the sign of
	the third component of the spin for one of the bosons coincides with the helicity 
	while for the other one with minus the helicity.}
\begin{multline}
	\frac{1}{\sigma} \frac{d\sigma}{dm_1\, dm_2}(m_1, m_2, c) \\
	= N \frac{\lambda^{1/2}(M^2,m_1^2,m_2^2) m_1^3 m_2^3}{D(m_1) D(m_2)}
	\tr[\tilde{\rho}(m_1,m_2,c)],
	\label{cross-sec-ZK}
\end{multline}
where $N$ is a normalization factor independent of $m_1$, $m_2$ and $c$; 
while the functions $\lambda$ and $D$ are defined 
in Eqs.~(\ref{lambda-def},\ref{D-def}).
Comparing (\ref{cross-sec-masses}) with (\ref{cross-sec-ZK}) (and taking into account
(\ref{rho-non-normalized}))
we finally obtain the probability distribution
\begin{equation}
	\mathcal{P}_{c}(m_1,m_2) 
	= N \frac{\lambda^{1/2}(M^2,m_1^2,m_2^2) m_1^3 m_2^3}{D(m_1) D(m_2)}
	\big[2+\kappa^2\big].
	\label{probability-distribution-final}
\end{equation} 
For a given value of $c$ the normalization factor $N$ can be determined numerically.
In \cite{ASBCM2022_entanglement_HtoZZ} the probability distribution 
$\mathcal{P}_{c=0}(m_1,m_2)$ has been obtained with the help of Monte Carlo
simulation. The results coincide with those computed from
(\ref{probability-distribution-final}).

Therefore, using (\ref{kappa},\ref{rho_ZZ_mixed-def},\ref{probability-distribution-final}), 
the density matrix averaged over kinematical configurations can be written as
\begin{equation}
\rho_{ZZ}(c) = 
\frac{1}{2a+b}
\begin{pmatrix}
0 & 0 & 0 & 0 & 0 & 0 & 0 & 0 & 0\\
0 & 0 & 0 & 0 & 0 & 0 & 0 & 0 & 0\\
0 & 0 & \fbox{$a$} & 0 & \fbox{$-d$} & 0 & \fbox{$a$} & 0 & 0\\
0 & 0 & 0 & 0 & 0 & 0 & 0 & 0 & 0\\
0 & 0 & \fbox{$-d$} & 0 & \fbox{$b$} & 0 & \fbox{$-d$} & 0 & 0\\
0 & 0 & 0& 0 & 0 & 0 & 0 & 0 & 0\\
0 & 0 & \fbox{$a$} & 0 & \fbox{$-d$} & 0 & \fbox{$a$} & 0 & 0\\
0 & 0 & 0 & 0 & 0 & 0 & 0 & 0 & 0\\
0 & 0 & 0 & 0 & 0 & 0 & 0 & 0 & 0
\end{pmatrix},
\label{rhoZZ-mixed-final}
\end{equation}
where for better visibility we have framed the non-zero matrix elements:
\begin{align}
a & = \fb{0},\\
b & = \fb{2} + 2 c \big[\fb{2}-\fb{0}\big] \nonumber\\
 &\phantom{=\,}+ c^2 \big[\fb{2}+\fb{-2}-2\fb{0}\big],\\
d & = \fb{1} + c \big[ \fb{1} - \fb{-1} \big],
\end{align}
and we have introduced the following notation
\begin{equation}
\fb{n} = \int\limits_{0\le m_1+m_2 \le M} \!\!\! dm_1 dm_2 \frac{\lambda^{1/2}(M^2,m_1^2,m_2^2) m_1^3    
	m_2^3}{D(m_1) D(m_2)} \beta^n,
\end{equation}
for $n=-2,-1,0,1,2$.

Note that sometimes it is relevant for phenomenological purposes to implement 
cuts on the possible values of the boson masses 
(for example to remove part of the background of a certain scattering process).
This feature is easily implemented theoretically via the integrals 
defining $\fb{n}$.
For instance, when considering a lower cut in the off-shell mass of the vector 
boson, $m_i \ge m^{cut}_i$, one
just needs to modify the lower bound in the integral over the corresponding mass:
\begin{multline}
\int\limits_{0\le m_1+m_2 \le M} dm_1 dm_2 = \int_{0}^{M} dm_1
\int_{0}^{M-m_1} dm_2\\
\to \int_{m^{cut}_1}^{M} dm_1
\int_{m^{cut}_2}^{M-m_1} dm_2.
\end{multline}

When we insert the measured values for the Higgs mass, $Z$ mass and $Z$ decay width,
i.e., $M=125.25\; \mathsf{GeV}$, $m_Z=91.19\; \mathsf{GeV}$,
$\Gamma_Z=2.50\; \mathsf{GeV}$ \cite{Workman:2022-PDG} we obtain
\begin{align}
a_Z & = 2989.76, \label{a-experimental}\\
b_Z & = 9431.55 + 12883.6 c + 4983.07 c^2, \label{b-experimental}\\
d_Z & = 4819.07 + 2752.19 c. \label{d-experimental}
\end{align}

\section{Bell inequalities and entanglement}

Now, we are at a position to discuss the violation of Bell inequalities in a system
of two $ZZ$ bosons. 
Various Bell inequalities have been designed for detecting departures from local realism
by quantum mechanical systems \cite{BCPSW2014}, the most popular one being
the CHSH inequality \cite{cab_CHSH1969}.
For a system consisting of two $d$-dimensional subsystems the optimal Bell inequality 
was formulated in \cite{CGLMP_Bell_ineq_high_spin,KKCZ_PhysRevA.65.032118} 
and is known as the CGLMP inequality.
For two qubits it reduces to the CHSH inequality.
We consider here two spin-1 particles therefore we present
the CGLMP inequality for $d=3$. 
We assume that Alice (Bob) can perform 
two possible measurements $A_1$ or $A_2$ ($B_1$ or $B_2$) on her (his) subsystem,
respectively. 
Each of these measurements can have three outcomes: 0,1,2.
Let $P(A_i=B_j+k)$ denotes the probability that the outcomes $A_i$ and $B_j$ differ
by $k$ modulo 3, 
i.e., $P(A_i=B_j+k) = \sum_{l=0}^{l=2} P(A_i=l,B_j=l+k \mod 3)$,
and let us define the following quantity
\begin{multline}
	\mathcal{I}_3 = \big[
	P(A_1=B_1) + P(B_1=A_2+1)\\
	+ P(A_2=B_2) +P(B_2=A_1)
	\big]\\
	-\big[
	P(A_1=B_1-1) + P(B_1=A_2)\\
	+P(A_2=B_2-1) + P(B_2=A_1-1)
	\big].
	\label{I3-def}
\end{multline}
The CGLMP inequality has the form
\begin{equation}
	\mathcal{I}_3 \le 2.
	\label{CGLMP-def}
\end{equation}
We assume that Alice can perform measurements on one of the $Z$ bosons, 
Bob on the second one.
In principle, what they can measure are spin projections on given directions.
A few remarks are in order here.
In \cite{BCR2022-Bell-vector-bosons} we discussed broader the problem
of choice of the proper spin observable. We have advocated there 
the Newton--Wigner spin operator. Under our assumptions, 
when Alice applies this spin operator to the basis vector $\ket{\lambda,\sigma}$,
this action can be written as the action of 
$(\Vec{S}_{\lambda^\prime\lambda}\otimes I)$
on $\ket{\lambda^\prime}\otimes\ket{\sigma}$, where 
$S^i$ are standard spin-1 matrices (and analogously for Bob).
Therefore, from now on we take Alice (Bob) observables as $A\otimes I$
($I\otimes B$) and identify
$\ket{\lambda}\otimes\ket{\sigma}\equiv\ket{\lambda,\sigma}$.

\subsection{Probabilities of spin projection measurements for a particular configuration}

In general, when discussing the violation of CGLMP inequality by the state
$\rho_{ZZ}(c)$ (Eq.~(\ref{rhoZZ-mixed-final})) we do not restrict
our attention to spin projection measurements only.
However, for a particular configuration, the probabilities $P_{\lambda\sigma}$
that Alice and Bob receive $\lambda$ and $\sigma$ when they measure spin
projections along directions $\vec{a}$ and $\vec{b}$ 
in the pure state (\ref{scalar-ZZ-n}), respectively, can be calculated explicitly.
We present them below
\begin{multline}
	P_{\pm\pm}  =  \frac{1}{4[2+\kappa^2]}
	\Big\{
	\big[1-(\vec{a}\cdot\vec{b})\big]^2 \\
	+ 2(\kappa-1) \big[ 1 - (\vec{a}\cdot\vec{b}) + \big( 1+(\vec{a}\cdot\vec{b}) \big)
	(\vec{a}\cdot\vec{n})(\vec{b}\cdot\vec{n})
	-(\vec{a}\cdot\vec{n})^2\\
	 - (\vec{b}\cdot\vec{n})^2 \big]
	+ (\kappa-1)^2 \big[1-(\vec{a}\cdot\vec{n})^2\big] 
	\big[1-(\vec{b}\cdot\vec{n})^2 \big]
	\Big\},
\end{multline}
\begin{multline}
	P_{\pm\mp}  =  \frac{1}{4[2+\kappa^2]}
	\Big\{
	\big[1+(\vec{a}\cdot\vec{b})\big]^2 \\
	+ 2(\kappa-1) \big[ 1 + (\vec{a}\cdot\vec{b}) - \big( 1-(\vec{a}\cdot\vec{b}) \big)
	(\vec{a}\cdot\vec{n})(\vec{b}\cdot\vec{n})
	-(\vec{a}\cdot\vec{n})^2\\
	- (\vec{b}\cdot\vec{n})^2 \big]
	+ (\kappa-1)^2 \big[1-(\vec{a}\cdot\vec{n})^2\big] \big[1-(\vec{b}\cdot\vec{n})^2 \big]
	\Big\},
\end{multline}
\begin{multline}
	P_{0\pm}  =  \frac{1}{2[2+\kappa^2]}
	\Big\{ 
	1 + (\kappa^2-1)(\vec{a}\cdot\vec{n})^2  \\
	- \big[ (\vec{a}\cdot\vec{b})
	+(\kappa-1)(\vec{a}\cdot\vec{n}) (\vec{b}\cdot\vec{n})\big]^2
	\Big\},
\end{multline}
\begin{multline}
	P_{\pm0}  =  \frac{1}{2[2+\kappa^2]}
	\Big\{ 
	1 + (\kappa^2-1)(\vec{b}\cdot\vec{n})^2  \\
	- \big[ (\vec{a}\cdot\vec{b})+(\kappa-1)(\vec{a}\cdot\vec{n}) (\vec{b}\cdot\vec{n})\big]^2
	\Big\},
\end{multline}
\begin{multline}
	P_{00}  =  \frac{1}{2+\kappa^2}
	\big[ (\vec{a}\cdot\vec{b})+(\kappa-1)(\vec{a}\cdot\vec{n}) (\vec{b}\cdot\vec{n})\big]^2.
\end{multline}
For the case $m_1=m_2$ the above probabilities coincide with the probabilities 
found in \cite{BCR2022-Bell-vector-bosons}.

It is worth noticing that if Alice and Bob are allowed to use only spin projections
as observables then,
in principle, the violation of the Bell inequality would be suboptimal as we are not covering 
the whole space of possible observables.

\subsection{Bell inequalities in a general ZZ state}

Now, we want to answer the question whether the state (\ref{rhoZZ-mixed-final})
violates the CGLMP inequality. In general, for a given state $\rho$ there does not
exist a simple way to find optimal observables $A_1$, $A_2$, $B_1$, $B_2$,
i.e., such observables for which the value of $\mathcal{I}_3$ is maximal in the
state $\rho$.
Usually, optimal observables are looked for with the help of a certain optimization
procedure. 
In \cite{ASBCM2022_entanglement_HtoZZ} such a procedure was proposed in the
considered there Standard Model coupling case, i.e. for $c=0$.
In this procedure one modifies the well known optimal choice of observables
for the maximally entangled state.
For completeness we describe the details of
this procedure in \ref{sec:AppB-sub1}.
This procedure works very well  for the case $c=0$ and for $c$ close to that value.
However, for higher values of $|c|$ this procedure gives the observables which
do not violate the CGLMP inequality.

Thus, we have considered also a different optimization procedure.
This procedure is inspired by the proof of Theorem 2 
in \cite{PR1992-GenericQNonlocality}. The details of this approach we described
in \ref{sec:AppB-sub2}.

\begin{figure}
\includegraphics[width=0.99\columnwidth]{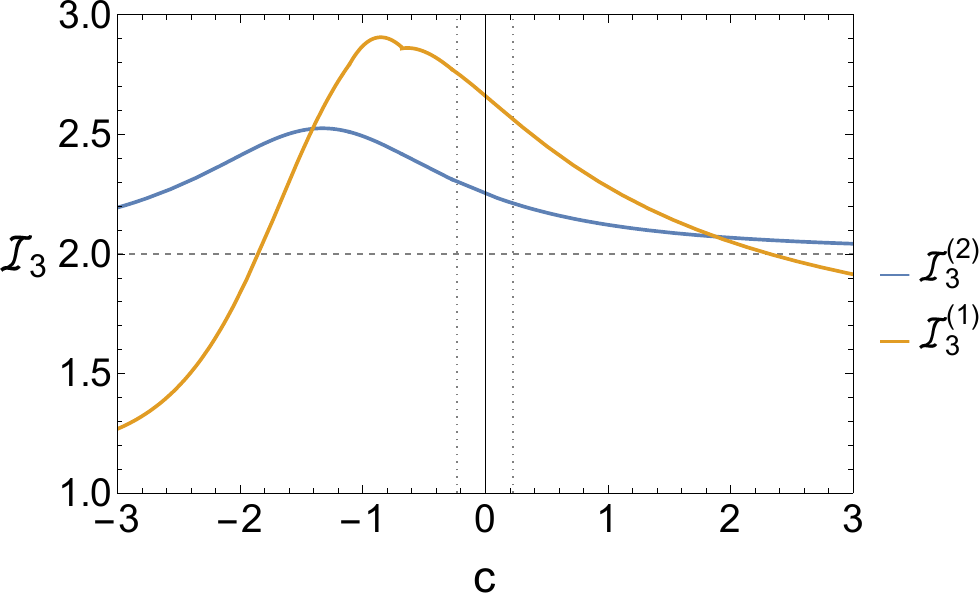}
\caption{In this figure we present the maximal value of $\mathcal{I}_3$
in the state (\ref{rhoZZ-mixed-final}) as a function of $c$. We have inserted
the measured values for the Higgs
mass, Z mass and Z decay width, i.e., we put $a$, $b$, $d$ given in 
Eqs.~(\ref{a-experimental},\ref{b-experimental},\ref{d-experimental}).
The $\mathcal{I}_{3}^{(1)}$ curve was obtained with the help of the optimization
procedure described in  \ref{sec:AppB-sub1} while the curve
$\mathcal{I}_{3}^{(2)}$ with the help of the procedure from 
\ref{sec:AppB-sub2}.
Vertical dotted lines delimit the range $(-c_{\mathsf{HZZ}}^{\mathsf{max}},c_{\mathsf{HZZ}}^{\mathsf{max}})$
(with $c_{\mathsf{HZZ}}^{\mathsf{max}}=0.26$---compare Eq.~(\ref{cHZZmax}))
of the parameter $c$ admissible by experimental data for the process $H\to ZZ$.}
\label{Fig1}	
\end{figure}

In Fig.~\ref{Fig1} we present the maximal value of $\mathcal{I}_3$
as a function of $c$ obtained with the help of 
both mentioned above optimization strategies.
This plot shows 
that the state $\rho_{ZZ}(c)$ can violate the CGLMP inequality
(\ref{rhoZZ-mixed-final}) for all values of $c$.
For $c\in (c_{-}^{Z},c_{+}^{Z})$,
where $c_{-}^{Z}=-1.3749$, $c_{+}^{Z}=1.6690$,
the optimization procedure proposed in \cite{ASBCM2022_entanglement_HtoZZ}
gives higher violation of CGLMP inequality than the procedure proposed
in \ref{sec:AppB-sub2}.
For other values of $c$ the situation is opposite. 
The highest value of $\mathcal{I}_3$ we obtained is equal to $2.9047$,
it is attained for $c_{\mathsf{max}}^{Z}=-0.8536$.

Regarding the values $|c| < c_{\mathsf{HZZ}}^{\mathsf{max}}$, 
larger violation of order $[2.5,2.8]$ is obtained when the optimization strategy 
from \ref{sec:AppB-sub1} is implemented, while violations
of order $[2.2,2.3]$ are attained for the optimization presented in \ref{sec:AppB-sub2}.

\subsection{Entanglement of a general $ZZ$ state}

To evaluate entanglement of the state (\ref{rhoZZ-mixed-final}) we use the
computable entanglement measure called logarithmic negativity
\cite{VW2002,Plenio_Logarithmic-negativity}
\begin{equation}
E_N(\rho) = \log_3(||\rho^{T_B}||_1),
\end{equation}
where $||A||_1=\tr(\sqrt{A^\dagger A})$ is the trace norm of a matrix $A$
and $T_B$ denotes partial transposition with respect to the second subsystem.
The trace norm of a matrix $A$ is equal to the sum of all the singular values of $A$;
when $A$ is hermitian then $||A||_1$ is equal to the sum of absolute values of all
eigenvalues of $A$.

\begin{figure}
\includegraphics[width=0.99\columnwidth]{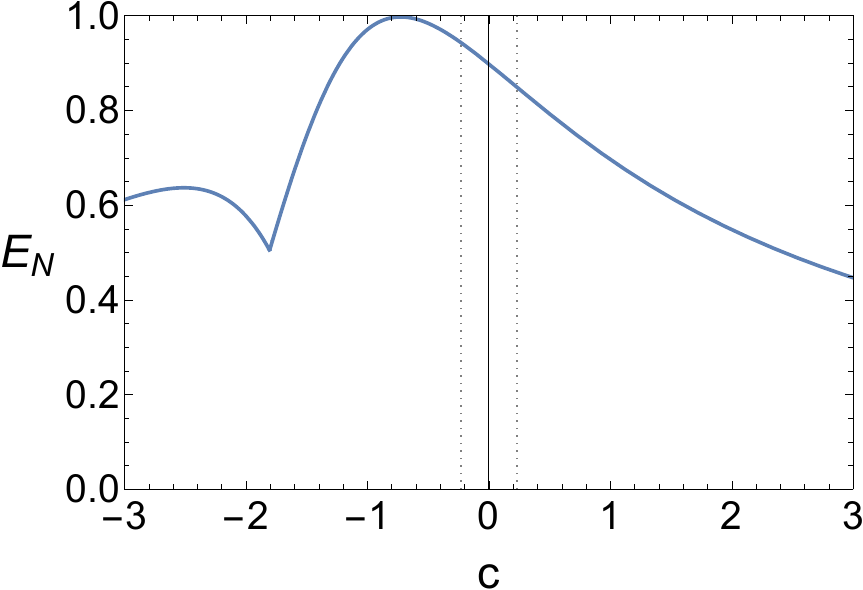}
\caption{In this figure we present  logarithmic negativity of the state
(\ref{rhoZZ-mixed-final}), $E_N(\rho_{ZZ}(c))$ as a function of $c$.
We have inserted the measured values for the Higgs
mass, Z mass and Z decay width, i.e., we put $a$, $b$, $d$ given in 
Eqs.~(\ref{a-experimental},\ref{b-experimental},\ref{d-experimental}).
The  cusp occurs for the value of $c$ for which the parameter $d_Z$ in the density 
matrix vanishes and is caused by the change in monotonicity of the trace norm of 
the partially transposed matrix.
Vertical dotted lines delimit the range $(-c_{\mathsf{HZZ}}^{\mathsf{max}},c_{\mathsf{HZZ}}^{\mathsf{max}})$
	(with $c_{\mathsf{HZZ}}^{\mathsf{max}}=0.26$---compare Eq.~(\ref{cHZZmax}))
	of the parameter $c$ admissible by experimental data for the process $H\to ZZ$.}
\label{Fig2}	
\end{figure}

If a state $\rho$ is separable then the logarithmic negativity of $\rho$ is equal to zero.
Thus, $E_N(\rho)>0$ indicates that the state $\rho$ is entangled.
In Fig.~\ref{Fig2} we have plotted the logarithmic negativity of the state
(\ref{rhoZZ-mixed-final}) with 
$a_Z$, $b_Z$, and $d_Z$
given in
Eqs.~(\ref{a-experimental},\ref{b-experimental},\ref{d-experimental}).
We see that the state is entangled for all values of $c$,
the maximal value of the logarithmic negativity equal to $0.9964$
is attained for $c=-0.7371$.
It is worth noticing that the state with the highest entanglement corresponds to
$c=-0.7371$ while the state with the highest violation of the CGLMP inequality
corresponds to $c=-0.8536$, i.e., these states are different.
This observation is consistent with the general property of CGLMP inequality
\cite{ADGL2002_PhysRevA.65.052325}.

Concerning the logarithmic negativity for 
$|c| < c_{\mathsf{HZZ}}^{\mathsf{max}}$,
it this is considerably far apart from 0, within a range $\sim [0.85, 0.95]$. 
This indicates a high
grade of entanglement in any $ZZ$ pair stemming from Higgs decays.

\subsection{Impact of a background}

Because the reconstruction of $\rho_{ZZ}(c)$ in a collider experiment is done 
via quantum tomography methods \cite{ASBCM2022_entanglement_HtoZZ,APBW2022_entanglement-weak-decays,Bernal2023-tomography}, 
the presence of systematic and statistical errors as well as the existence of a small
background in 
$H\to ZZ \to f_1^+ f_1^- f_2^+ f_2^-$
processes lead to a modification of its exact form.
Following the discussion given in 
\cite{ASBCM2022_entanglement_HtoZZ}, we will focus on analyzing 
$H\to ZZ \to e^+ e^- \mu^+ \mu^-$, since it
constitutes one of the cleanest channels to be explored at the LHC. 
In principle, the two $Z$ bosons are cleanly identified. 
Due to the fact that one of them is nearly on-shell, it gives two leptons whose
invariant mass is close to $m_Z$, while the remaining two leptons have a much 
lower invariant mass.
We have labeled the (very close to real) $Z$ boson with largest invariant mass as 
$Z_1$ and its four-momentum could be reconstructed from its decay products
$l_1^+ l_1^-$. On the other hand, the off-shell
$Z$ boson is labeled as $Z_2$ and its momentum is determined summing up the 
momenta of its decay products $l_2^+ l_2^-$.

\begin{figure}
\includegraphics[width=0.99\columnwidth]{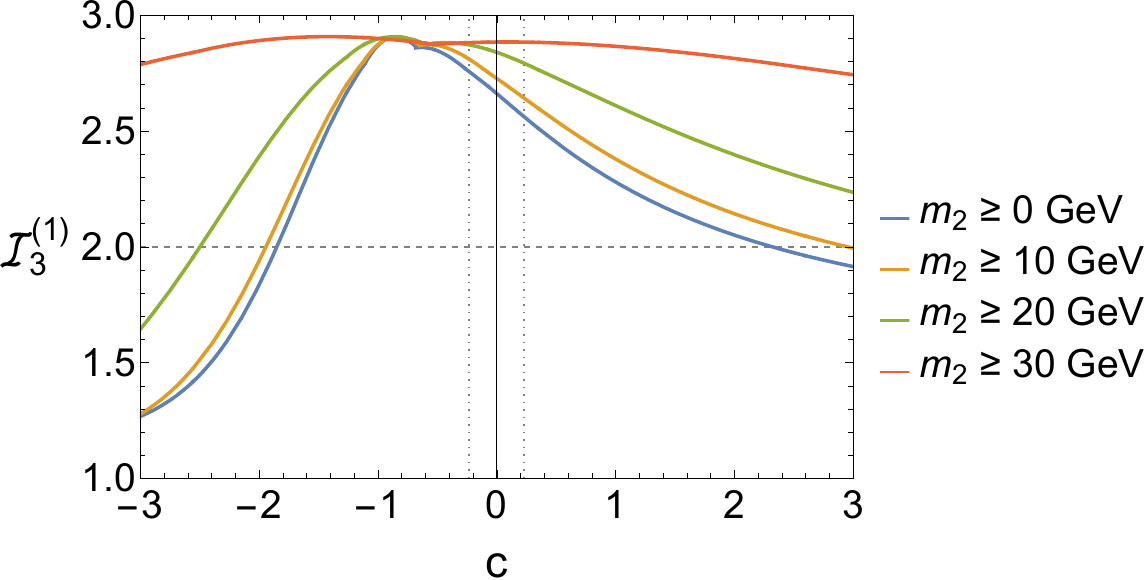}
\caption{In this figure we present the maximal value of $\mathcal{I}_3$
in the state (\ref{rhoZZ-mixed-final}) as a function of $c$. We have inserted
the measured values for the Higgs
mass, Z mass and Z decay width, i.e., we put $a$, $b$, $d$ given in 
Eqs.~(\ref{a-experimental},\ref{b-experimental},\ref{d-experimental}).
We have applied the optimization
procedure described in \ref{sec:AppB-sub1} 
and assumed the 
cuts $m_2 \ge 0$, $10$, $20$, $30\; \mathsf{GeV}$ are implemented.
Vertical dotted lines delimit the range $(-c_{\mathsf{HZZ}}^{\mathsf{max}},c_{\mathsf{HZZ}}^{\mathsf{max}})$
	(with $c_{\mathsf{HZZ}}^{\mathsf{max}}=0.26$---compare Eq.~(\ref{cHZZmax}))
	of the parameter $c$ admissible by experimental data for the process $H\to ZZ$.}
\label{Fig3}	
\end{figure}

\begin{figure}
\includegraphics[width=0.99\columnwidth]{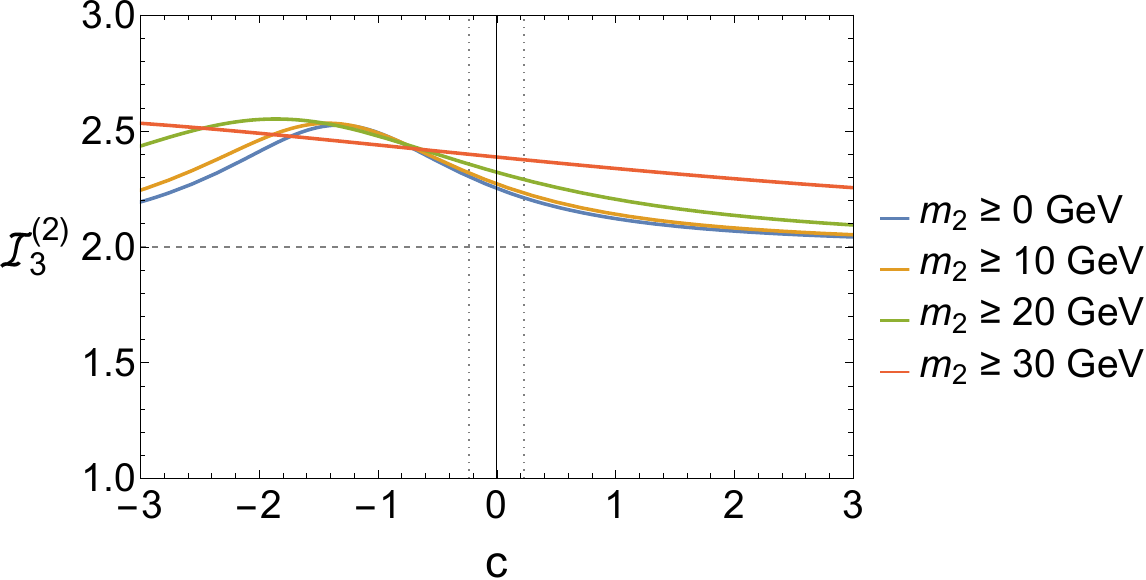}
\caption{In this figure we present the maximal value of $\mathcal{I}_3$
in the state (\ref{rhoZZ-mixed-final}) as a function of $c$. We have inserted
the measured values for the Higgs
mass, Z mass and Z decay width, i.e., we put $a$, $b$, $d$ given in 
Eqs.~(\ref{a-experimental},\ref{b-experimental},\ref{d-experimental}).
We have applied the optimization
procedure described in \ref{sec:AppB-sub2} 
and assumed the 
cuts $m_2 \ge 0$, $10$, $20$, $30\; \mathsf{GeV}$ are implemented.
Vertical dotted lines delimit the range $(-c_{\mathsf{HZZ}}^{\mathsf{max}},c_{\mathsf{HZZ}}^{\mathsf{max}})$
	(with $c_{\mathsf{HZZ}}^{\mathsf{max}}=0.26$---compare Eq.~(\ref{cHZZmax}))
	of the parameter $c$ admissible by experimental data for the process $H\to ZZ$.}
\label{Fig4}	
\end{figure}

\begin{figure}
\includegraphics[width=0.99\columnwidth]{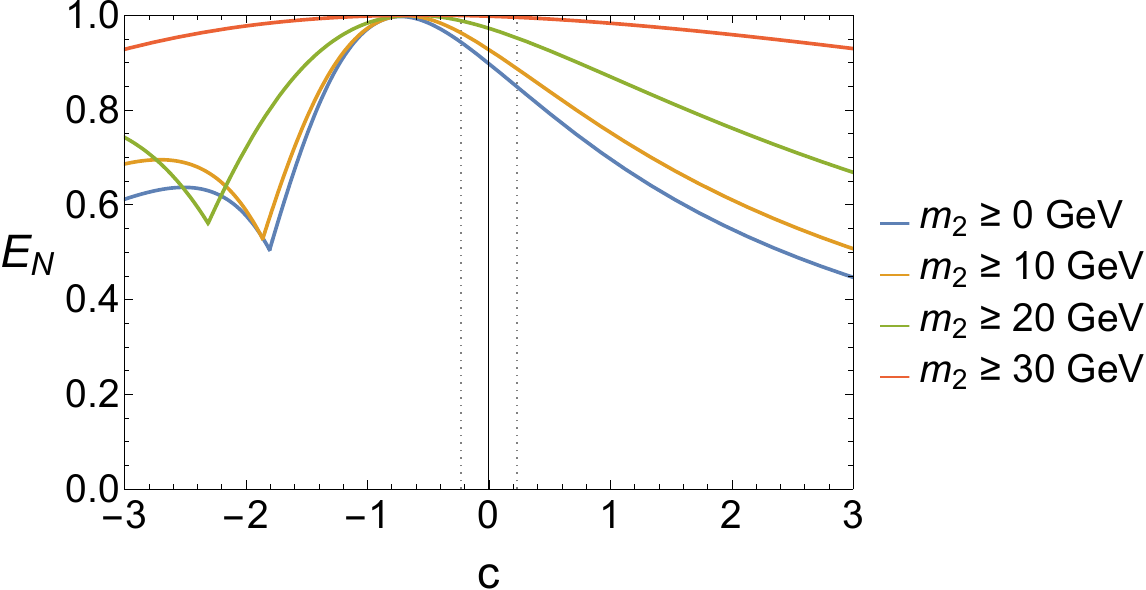}
\caption{In this figure we present  logarithmic negativity of the state
(\ref{rhoZZ-mixed-final}), $E_N(\rho_{ZZ}(c))$ as a function of $c$.
We have inserted the measured values for the Higgs
mass, Z mass and Z decay width, i.e., we put $a$, $b$, $d$ given in 
Eqs.~(\ref{a-experimental},\ref{b-experimental},\ref{d-experimental}).
We have assumed the 
cuts $m_2 \ge 0$, $10$, $20$, $30\; \mathsf{GeV}$ are implemented.
Vertical dotted lines delimit the range $(-c_{\mathsf{HZZ}}^{\mathsf{max}},c_{\mathsf{HZZ}}^{\mathsf{max}})$
	(with $c_{\mathsf{HZZ}}^{\mathsf{max}}=0.26$---compare Eq.~(\ref{cHZZmax}))
	of the parameter $c$ admissible by experimental data for the process $H\to ZZ$.}
\label{Fig5}	
\end{figure}

Concerning both the systematic and statistical errors of the tomography procedure, 
it was stated in \cite{ASBCM2022_entanglement_HtoZZ}
that the statistical one dominates with respect to the systematics and detector resolution. 
An estimation of the former was computed in this previous paper and the results show 
that, even with these errors, a violation of Bell inequalities for the $\rho_{ZZ}(0)$ 
state could be probed at the $4.5\sigma$ level in the HL-LHC. 
Regarding the background of the process, the main one comes from the electroweak
one $pp \to ZZ/Z\gamma  \to 4 l$, 
being this one about 4 times smaller at the Higgs peak \cite{CMSCollab2021-H-to-4-leptons}.
Nevertheless,
as claimed in \cite{ASBCM2022_entanglement_HtoZZ}, a background subtraction 
will be necessary before computing the entanglement
observable and evaluating the CGLMP inequality. In general, the non-negligible background will
slightly contribute to the statistical uncertainty of the measurements.
Moreover, as proposed in \cite{ASBCM2022_entanglement_HtoZZ}, 
the larger the invariant mass $m_2$ of the off-shell $Z$ boson, the more
entangled the $\rho_{ZZ}(c)$ state. 
Therefore, requiring a lower cut on $m_2$ leads to an interplay between
increasing the entanglement (hence the violation of the CGLMP inequality) and decreasing
the statistics (thus increasing the uncertainty in the measurements). 
In Figs.~\ref{Fig3}, \ref{Fig4}, \ref{Fig5} we have studied 
the theoretical dependence of $\mathcal{I}_3$ and $E_N$
on $c$, once the 
cuts $m_2 \ge 0, 10, 20, 30\; \mathsf{GeV}$ are implemented.

Returning for a moment to our discussion of ``offshellness'' of $Z$ bosons
produced in the decay (\ref{HZZ}), let us notice that $\mathcal{I}_{3}^{(1)}$
and $\mathcal{I}_{3}^{(2)}$ obtained for $m_1=m_Z$, 
$0\;\mathsf{GeV}\le m_2 \le m_H-m_Z$ (blue curves in Figs.~\ref{Fig3} and
\ref{Fig4}) are identical as $\mathcal{I}_{3}^{(1)}$
and $\mathcal{I}_{3}^{(2)}$ plotted in Fig.~\ref{Fig1} where we allowed arbitrary
masses $m_1$ and $m_2$ (of course constrained by the four-momentum conservation).
The same holds for $E_N$ plotted in Figs.~\ref{Fig5} (blue line) and \ref{Fig2}.
Of course this coincidence is not accidental, it results from the fact that
the probability distribution $\mathcal{P}_c(m_1,m_2)$ is peaked at $m_i=m_Z$.

\begin{figure}
\includegraphics[width=0.99\columnwidth]{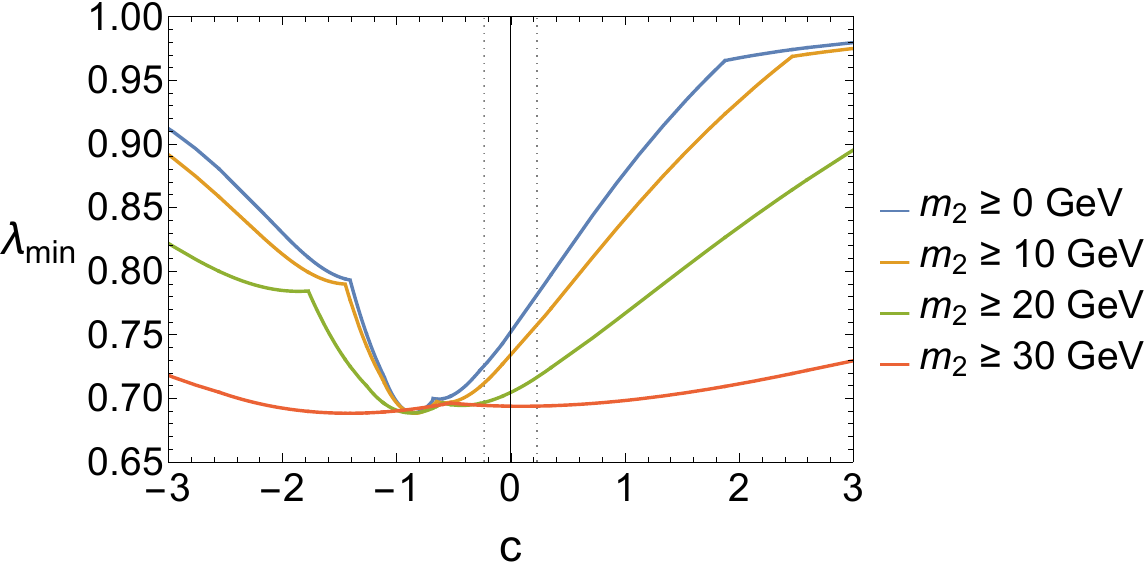}
\caption{In this figure we present  
$\lambda_{\mathsf{min}}$
(\ref{lambda_min_final}), as a function of $c$.
We have inserted the measured values for the Higgs
mass, Z mass and Z decay width, i.e., we put $a$, $b$, $d$ given in 
Eqs.~(\ref{a-experimental},\ref{b-experimental},\ref{d-experimental}).
We have assumed the 
cuts $m_2 \ge 0$, $10$, $20$, $30\; \mathsf{GeV}$ are implemented.
Vertical dotted lines delimit the range $(-c_{\mathsf{HZZ}}^{\mathsf{max}},c_{\mathsf{HZZ}}^{\mathsf{max}})$
	(with $c_{\mathsf{HZZ}}^{\mathsf{max}}=0.26$---compare Eq.~(\ref{cHZZmax}))
	of the parameter $c$ admissible by experimental data for the process $H\to ZZ$.}
\label{Fig6}	
\end{figure}

Finally, in order to estimate the overall allowed uncertainty in both the entanglement 
and the Bell inequality violation, we have estimated the noise resistance of the CGLMP
inequality violation with respect to the white noise. To this aim we considered the
state $\rho_{ZZ}(c)$ mixed with the identity operator, i.e. the state
\begin{equation}
\lambda \rho_{ZZ}(c) + (1-\lambda) \tfrac{1}{9} I_9,\qquad\lambda \in (0, 1].
\label{rho_ZZ-with-noise}
\end{equation}
Now, the noise resistance we define as a minimal value of $\lambda$, 
$\lambda_{\mathsf{min}}$, for which the state (\ref{rho_ZZ-with-noise}) violates
the CGLMP inequality. Inserting the state (\ref{rho_ZZ-with-noise}) into the
CGLMP inequality (\ref{CGLMP-OBell}) and taking into account that
$\tr(\mathcal{O}_{\mathsf{Bell}})=0$ we obtain
\begin{equation}
\lambda_{\mathsf{min}} = \frac{2}{\max\{
\tr(\rho_{ZZ}(c)\mathcal{O}_{\mathsf{Bell}}) \}}.
\end{equation}
We obtained the maximal value of 
$\tr(\rho_{ZZ}(c)\mathcal{O}_{\mathsf{Bell}})=\mathcal{I}_3$
with the help of two different optimization procedures ( \ref{sec:AppB-sub1}
and \ref{sec:AppB-sub2}) and denoted as $\mathcal{I}_{3}^{(1)}$
and $\mathcal{I}_{3}^{(2)}$, respectively (compare Fig.~\ref{Fig1}).
Therefore, we have
\begin{equation}
\lambda_{\mathsf{min}} = \frac{2}{\max\{ \mathcal{I}_{3}^{(1)}, \mathcal{I}_{3}^{(2)}\}}
\label{lambda_min_final}
\end{equation}
and this value we have plotted in Fig.~\ref{Fig6}.

The plots show that for values of $c$ close to 0 (which are the expected ones, 
due to the present bounds in anomalous couplings for the $HZZ$ vertex 
\cite{RRS2021-anomalous-gauge-H}), 
one can stand up to almost a 20\% of noise and still attain a violation of the CGLMP inequality and hence an entangled state. 
Actually, the resistance to noise increases with the invariant mass of the off-shell $Z$ 
boson, reaching for the cut $m_2 \ge 30\; \mathsf{GeV}$ a resistance of 20\% 
for $c \in [-3, 3]$ and a resistance of almost 30\% for $c \in [-2, 2]$.

\section{Conclusions}

In conclusions, we have analyzed entanglement and Bell inequality violation
in a system of two $Z$ bosons produced in Higgs decay. We consider beyond the
Standard Model structure of the vertex describing interaction of $H$ with daughter
bosons. The amplitude corresponding to the most general Lorentz-invariant, 
CPT conserving coupling of a (pseudo)scalar particle with two vector bosons
depends on three coupling constants $v_1$, $v_2$, $v_3$,
and is explicitly given in Eq.~(\ref{general-vertex}).
The Standard Model interaction corresponds to $v_1=1$, $v_2=v_3=0$
while $v_3\not=0$ implies Higgs boson with a pseudo-scalar
component and indicates the possibility of CP violation.
In this paper we have considered the case $v_3=0$, $v_1\not=0$, $v_2$ free,
i.e. we have assumed the scalar Higgs boson but admitted a beyond Standard 
Model coupling $v_2\not=0$. In such a case, the state of produced bosons,
beyond four-momenta and spins, can be
characterized by a single parameter $c$ which, up to normalization is equal
to $v_2/v_1$ (compare Eq.~(\ref{c-v_2-v_1})).
Under such assumptions, in the center-of-mass frame, we have determined 
the most general state of $ZZ$ boson pair for a particular event $H\to ZZ$.
Next, we have considered a more realistic case when data are collected from
different kinematical configurations. In such a situation a $ZZ$ state can be calculated
by averaging over those configurations with respect to a proper probability distribution 
function (PDF). 
Thus, assuming further that each $Z$ boson decays into fermion-antifermion pair,
we have derived the corresponding PDF and computed the $ZZ$ boson density matrix. 
Finally, we have shown that this matrix is entangled and violates the CGLMP
inequality for all values of coupling (i.e. for all values of $c$)
including the range admissible by experimental data
\cite{CMSCollab2019-H-anomalous-PhysRevD.99.112003}.
The procedure to check this is completely general and can be applied for any 
other decay of a scalar particle into vector bosons, with their corresponding 
decay to fermions, once the PDF of the latter decay is known. 

Moreover, our preliminary studies show that
the inclusion of a CP-odd anomalous coupling should not 
qualitatively change the results derived. 
However, in this case the optimization strategy is more involved
and work is still in progress.	

Summarizing, this work settles a constructive way of 
probing the entanglement and violation of Bell
inequalities of any vector boson pair coming from a spin-0 particle, independently 
of the value of the couplings in hand (as long as the interactions among particles 
are CPT and Lorentz invariant). 
This feature is of a complete novelty in the literature and states the highly non-trivial fact 
that non locality (and hence entanglement) of vector bosons in these kinds of 
processes is theoretical ensured in any phenomenological model (it could have been 
the case in which although the state is entangled, it
does not violate any Bell inequality). 
Thus, the only limitation in checking the quantum behavior
of these processes comes from the experimental side. In particular, this work gives 
the theoretical framework to test the quantum nature of processes in a great variety 
of phenomenological models, covering for instance models with extended scalar sectors 
as well as axion like particles (ALP) models where we let the ALP to interact 
with vector bosons.


\begin{acknowledgements}
P.C. and J.R. are supported by the University of Lodz under IDUB project.
A.B. is grateful to J.A. Casas  and J. M. Moreno for very useful discussions. 
A.B. acknowledges the support of the Spanish Agencia Estatal de Investigacion 
through the grants ``IFT Centro de Excelencia Severo Ochoa CEX2020-001007-S''
and PID2019-110058GB-C22 funded by 
MCIN/AEI/10.13039/501100011033 and by ERDF. 
The work of A.B. is supported through the FPI grant PRE2020-095867 funded 
by MCIN/AEI/10.13039/501100011033. 
\end{acknowledgements}

\appendix
\section{Some definitions and useful formulas}

Following e.g. \cite{ZK_2016} we define the following functions
\begin{align}
\lambda(x,y,z) & = x^2+y^2+z^2 - 2xy - 2xz - 2yz,
\label{lambda-def}\\
D(m) & = \big(m^2-m_Z^2\big)^2 + (m_Z \Gamma_Z)^2,
\label{D-def}
\end{align}
where $m_Z, \Gamma_Z$ denotes the mass and decay width of the on-shell $Z$ boson.
In the CM frame the Higgs particle with the four-momentum $(M,\vec{0})$ decays into
two off-shell $Z$ bosons with four-momenta
$k^\mu=(\omega_1,\vec{k})$, $\omega_1^2 - \vec{k}^2 = m_1^2$ and
$p^\mu=(\omega_2,-\vec{k})$, $\omega_2^2 - \vec{k}^2 = m_2^2$.
The energy conservation gives
\begin{equation}
	M = \omega_1 +\omega_2.
	\label{energy_conservation}
\end{equation}
Using these equations in the CM frame we obtain
\begin{align}
\vec{k}^2 & = \frac{1}{4M^2} \lambda(M^2,m_1^2,m_2^2), 
\label{formula_1}\\
kp & = \frac{1}{2} \Big[M^2 - m_1^2 - m_2^2\Big], 
\label{formula_2}\\
\omega_1 & = \frac{1}{2M} \Big[M^2 + (m_1^2 - m_2^2)\Big], 
\label{formula_3}\\
\omega_2 & = \frac{1}{2M} \Big[M^2 - (m_1^2 - m_2^2)\Big].
\label{formula_4}
\end{align}

\section{Optimal observables for CGLMP violation}
\label{sec:AppB}

For completeness we describe here the optimization stra\-tegies used to obtain
observables $A_1$, $A_2$, $B_1$, 
and $B_2$ which maximize the value of $\mathcal{I}_3$ 
as depicted in Fig.~\ref{Fig1}.

The first procedure is similar to that 
applied in \cite{ASBCM2022_entanglement_HtoZZ}.
The second one is inspired by the proof of Theorem 2 in 
\cite{PR1992-GenericQNonlocality}.

First, it is clear that the CGLMP inequality (\ref{CGLMP-def}) can be written as
\begin{equation}
\tr\big( \rho \mathcal{O}_{\mathsf{Bell}}  \big) \le 2,
\label{CGLMP-OBell}
\end{equation}
where $\mathcal{O}_{\mathsf{Bell}}$ is a certain operator depending on the
observables $A_1$, $A_2$, $B_1$, and $B_2$.

Next, each hermitian $3\times3$ observable $A$ can be represented via 
the $3\times3$ unitary matrix $U_A$. This unitary matrix is defined in a simple way:
columns of $U_A$ are normalized eigenvectors of $A$ in a given basis.
With this notation one obtains \cite{ASBCM2022_entanglement_HtoZZ}
\begin{multline}
\mathcal{O}_{\mathsf{Bell}} = -[U_{A_1}\otimes U_{B_1}] P_1 [I\otimes S^3] 
P_{1}^{\dagger} [U_{A_1}\otimes U_{B_1}]^\dagger\\
+ [U_{A_1}\otimes U_{B_2}] P_0 [I\otimes S^3] 
P_{0}^{\dagger} [U_{A_1}\otimes U_{B_2}]^\dagger\\
+ [U_{A_2}\otimes U_{B_1}] P_1 [I\otimes S^3] 
P_{1}^{\dagger} [U_{A_2}\otimes U_{B_1}]^\dagger\\
- [U_{A_2}\otimes U_{B_2}] P_1 [I\otimes S^3] 
P_{1}^{\dagger} [U_{A_2}\otimes U_{B_2}]^\dagger,
\label{OBell-by-U}
\end{multline}
where $S^3$ is the standard spin $z$ component matrix, $S^3=\mathrm{diag}(1,0,-1)$,
and $P_0$, $P_1$ are $3^2\times 3^2$ block-diagonal  permutation matrices:
\begin{equation}
P_n = 
\begin{pmatrix}
C^n & \mathcal{O} & \mathcal{O} \\
\mathcal{O} & C^{n+1} & \mathcal{O} \\
\mathcal{O} & \mathcal{O} & C^{n+2}
\end{pmatrix}, \quad n=0,1, 
\end{equation}
where $\mathcal{O}$ is the $3\times 3$ null matrix and
$C$ is the $3\times3$ cyclic permutation matrix
\begin{equation}
C = 
\begin{pmatrix}
0 & 0 & 1\\
1 & 0 & 0\\
0 & 1 & 0	
\end{pmatrix}.
\end{equation}
Each $U$ from (\ref{OBell-by-U}) can be taken as an element of $SU(3)$ group,
this group has 8 parameters.
Thus, to perform the full optimization of $\mathcal{O}_{\mathsf{Bell}}$ for a given 
state one should check the $8^4$ dimensional parameter space.

\subsection{Strategy 1}
\label{sec:AppB-sub1}

To simplify this task, in \cite{ASBCM2022_entanglement_HtoZZ} the following 
approach was applied. It is known what is the form of the optimal Bell operator
for the maximally entangled state
$\rho_{ME}=\ket{\psi_{ME}}\bra{\psi_{ME}}$, 
$\ket{\psi_{ME}}=\tfrac{1}{\sqrt{3}}(\ket{++}+\ket{00}+\ket{--})$.
Let us denote this optimal Bell operator by $\mathcal{O}_{\mathsf{Bell}}^{ME}$.

For $\kappa=1$ the state $\ket{\psi_{ZZ}^{\mathsf{scalar}}}(m_1,m_2,c)$
(\ref{ZZ-state-CM}) reduces to 
\begin{equation}
\ket{\psi_{ZZ}^{\mathsf{scalar}}}|_{\kappa=1}=
\tfrac{1}{\sqrt{3}}(\ket{+-}-\ket{00}+\ket{-+}).
\end{equation}
Applying to $\ket{\psi_{ZZ}^{\mathsf{scalar}}}|_{\kappa=1}$
the operator $O_A\otimes I$, where
\begin{equation}
O_A = \begin{pmatrix}
0 & 0 & 1\\
0 & -1 & 0\\
1 & 0 & 0
\end{pmatrix},
\end{equation}
we obtain the maximally entangled state $\ket{\psi_{ME}}$.
Thus, the optimal Bell operator for the state
$\ket{\psi_{ZZ}^{\mathsf{scalar}}}|_{\kappa=1}$
has the form
\begin{equation}
(O_A\otimes I)^\dagger 
\mathcal{O}_{\mathsf{Bell}}^{ME}
(O_A\otimes I).
\label{OBell-OA}
\end{equation}
Next, we have
\begin{equation}
(U_A\otimes U_B) 
\sum_{\lambda\sigma} \Omega_{\lambda\sigma} 
\ket{\lambda,\sigma} 
= \sum_{\lambda\sigma} \Omega_{\lambda\sigma}^{\prime}
\ket{\lambda,\sigma},
\end{equation}
where
\begin{equation}
\Omega^\prime = U_A \Omega U_{B}^{T}.
\label{Omega-transf}
\end{equation}
For the maximally entangled state $\Omega^{ME}=\tfrac{1}{\sqrt{3}}I$,
thus this state is invariant on the action $U\otimes  U^*$.
Therefore, the optimal Bell observable for the state
$\ket{\psi_{ZZ}^{\mathsf{scalar}}}|_{\kappa=1}$,
instead of the form (\ref{OBell-OA}) can be written in an equivalent
form
\begin{equation}
(U O_A\otimes U^*)^\dagger 
\mathcal{O}_{\mathsf{Bell}}^{ME}
(U O_A\otimes U^*),
\label{OBell-OA-U}
\end{equation}
where $U$ is an arbitrary unitary matrix. 
The value of $\mathcal{I}_3$ 
with the Bell operator given in (\ref{OBell-OA-U})
in the state
$\ket{\psi_{ZZ}^{\mathsf{scalar}}}|_{\kappa=1}$ is the same
for all unitary matrices $U$.

For $\kappa\not=1$ the Bell operator (\ref{OBell-OA}) is no longer an optimal one.
Moreover, different choices of $U$ in (\ref{OBell-OA-U}) lead to different
values of $\mathcal{I}_3$ in this case. Thus, one can look for an optimal choice
taking the Bell operator in the form (\ref{OBell-OA-U}) and optimizing over 
all $U$ matrices. This can be simplified further by observing that the state
$(O_A\otimes I) \ket{\psi_{ZZ}^{\mathsf{scalar}}}(m_1, m_2, c)$, according to
(\ref{Omega-psiZZ},\ref{Omega-transf}) is represented by the matrix
\begin{equation}
\Omega_{O_A} = \frac{1}{\sqrt{2+\kappa^2}} 
\begin{pmatrix}
1 & 0 & 0\\
0 & \kappa & 0\\
0 & 0 & 1
\end{pmatrix},
\end{equation}
and that $\Omega_{O_A}$ is invariant under transformations (\ref{Omega-transf})
with
\begin{equation}
U_A= 
\begin{pmatrix}
\alpha & 0 & \beta\\
0 & e^{i\phi} & 0\\
\gamma & 0 & \delta
\end{pmatrix},\quad
U_B=U_A^*,\quad
\begin{pmatrix}
\alpha & \beta \\
\gamma & \delta
\end{pmatrix}\in SU(2).
\end{equation}
Therefore, optimization can be restricted to $U$ representing distinct cosets
of $U(3)/(SU(2)\times U(1))$.
In this paper, for the purpose of optimization,
we used the following parametrization of $U$
\begin{equation}
U = \exp(i \Theta),
\end{equation}
with
\begin{equation}
\Theta = 
\begin{pmatrix}
0 & e^{i\phi_1} R \cos\theta & 0\\
e^{-i\phi_1} R \cos\theta & 0 & e^{i\phi_2} R \sin\theta\\
0 & e^{-i\phi_2} R \sin\theta & 0
\end{pmatrix}.
\end{equation}
To obtain Fig.~\ref{Fig1}, for each value of $c$ we performed 
the optimization using the above parametrization.
The optimal choice of $R$, $\theta$, $\phi_1$, and $\phi_2$ depends on the value 
of $c$, this change of parametrization is responsible for the cusp appearing in the plot.

\subsection{Strategy 2}
\label{sec:AppB-sub2}

In this case we define a matrix
\begin{equation}
U_V(t) = 
\begin{pmatrix}
\cos\tfrac{t}{2} & 0 & \sin\tfrac{t}{2}\\
0 & 1 & 0\\
-\sin\tfrac{t}{2} & 0 & \cos\tfrac{t}{2}
\end{pmatrix}
\end{equation}
and assume that observables used by Alice and Bob are represented
by the following unitary matrices:
\begin{align}
U_{A_1} &= U_V(0),\quad
U_{A_2} = U_V(\tfrac{\pi}{2}),\\
U_{B_1} &=  U_V(t),\quad
U_{B_2} = U_V(-t).
\end{align}
Next, we are looking for such value of $t$ which gives the highest violation
of the CGLMP inequality. It appears that the optimal value is $t=-\tfrac{\pi}{4}$.
We used this value to plot Fig.~\ref{Fig1}.


%

\end{document}